\documentclass[titlepage,11pt]{article} 

\usepackage[utf8]{inputenc} 
\usepackage{verbatim}
\usepackage{amsmath}
\usepackage{amssymb}
\usepackage{graphicx}
\usepackage{lipsum}

\usepackage{geometry} 
\usepackage{graphicx} 

\usepackage{booktabs} 
\usepackage{array} 
\usepackage{paralist} 
\usepackage{verbatim} 
\usepackage{subfig} 

\usepackage{fancyhdr} 
\usepackage{sectsty}
\allsectionsfont{\sffamily\mdseries\upshape} 

\usepackage[nottoc,notlof,notlot]{tocbibind} 
\usepackage[titles,subfigure]{tocloft} 

\usepackage[nottoc]{tocbibind}

\usepackage[toc]{glossaries}

\title{Perimeter Scholars International Research Essay\\ {\ } \\ \bf{Scale Invariance in Gravity On The Light-Front} \\}
\author{Vasudev Shyam\\ {}  \\ \it{Under the supervision of} \\ {}\\ Lee Smolin \\ \\ \emph{Perimeter Institute for Theoretical Physics, Waterloo, ON N2L 2Y5, Canada} \\}

\begin{document}

\maketitle

\begin{abstract}

In general relativity, the double null foliation is one for which $d$-dimensional
spacetime is foliated by two families of intersecting null hyper surfaces
(i.e. surfaces whose normal vectors are null) of $(d-1)$ dimensions. Their intersection is at space
like surfaces of dimension $(d-2)$.
This means that the leaves of this foliation are the space like surfaces
of two dimensions lower than that of spacetime which are located by
looking at the bundles of light rays that are going into, and emanating
from them. Using this foliation, we present a reformulation of the
theory which makes explicit the true dynamical degrees of freedom.
This is accompanied by making manifest a hidden local conformal invariance.
Revealing this local symmetry comes at the cost of preferring a parameterisation
of the null hyper surfaces. More precisely, a preferred ruling of
the null surfaces by their generators needs to be chosen so that the theory whose gauge symmetries are enhanced by Weyl invariance restricted to preserve the foliation, is equivalent to general relativity.

I therefore find a dual theory that is locally equivalent to general
relativity but possessing enhanced local gauge symmetries: Weyl local
scale invariance and diffeomorphism invariance, both of which are
restricted to preserve the two families of null surfaces. This theory
is constructed employing the so called symmetry trading algorithm,
which shall be described in detail in this essay. Alternatively,
the theory can also be seen as a `phase' of a particular Scalar-Tensor
theory, because it is equivalent to a particular class of
configurations of a scalar field conformally coupled to general relativity.
\end{abstract}

\tableofcontents

\section{Introduction}

I explore a mechanism which allows one to modify gauge theories in
general and General Relativity in particular without introducing additional
propagating degrees of freedom. The modification has the effect of
altering the local gauge transformations the theory is invariant under,
so this mechanism is aptly dubbed``symmetry trading''.
This will result in identifying two theories, each of which can be
seen as a gauge fixing of the other. More precisely, the two theories
each possess a first class constraint\footnote{In the sense of Constrained Hamiltonian Dynamics which we shall briefly
review in the second part of this essay} such that one is the gauge fixing of the other. This mechanism was
first defined in (\cite{GGK}) in the context of Hamiltonian General
Relativity and through it, a theory known as Shape Dynamics was discovered.

This theory is one where the many fingered time re-foliation invariance
which is manifest in the Arnowitt-Deser-Misner formalism of Hamiltonian
General Relativity (\cite{ADM}) is gauge fixed by using a preferred
foliation. This is a foliation of spacetime by space-like hyper surfaces
of constant mean curvature (CMC). Then the theory is invariant under
foliation preserving diffeomorphisms and local Weyl transformations
that act only on the spatial hyper surfaces of the foliation as gauge
symmetries. So what is getting `traded' is the local re-foliation
invariance invariance of general relativity for the local spatial
Weyl invariance of Shape Dynamics. For an up to date review of the
theory of Shape Dynamics see (\cite{Mtut},\cite{GKFAQ}).

Shape Dynamics allows us to view the dynamics of general relativity
though the evolution of spatial three-geometries,\footnote{Here the use of the term `geometry' is to emphasise that only the
dynamics of the spatial metrics.modulo diffeomorphisms is of physical
relevance.} and this is dual to the conventional ADM formalism thanks to the
symmetry trading mechanism. It then behooves us to ponder how robust
this procedure is, and whether it could be applied to different approaches
to the dynamics of general relativity itself.

In particular, despite the many uses of the ADM formalism, it fails
to work in scenarios where the relevant dynamical hyper surfaces are
null. The primary difficulty in this setting comes from the degeneracy
of the three metric adapted onto the null surface which amongst other
things makes the embedding relations difficult to deal with. Nevertheless,
these difficulties that arise are most easily circumvented by employing
a foliation by co-dimension 2 space-like surfaces at the intersection
of two null hyper-surfaces each of co-dimension 1. This setting shall
be called the (2+2) double null embedding formalism.

The merits of this formalism include providing a natural setting for
studying black hole and apparent horizons (\cite{ISRHor}, \cite{HAYWHor}),
and is relevant in the context of light-front quantisation \cite{BUKHR}
and the Bousso bound (\cite{BOUS}). The application which shall be
of primary concern to us here will be its utility in the analysis
of the characteristic initial value problem for general relativity.
I will elucidate this in subsection 1.3 below. In short, this formalism
is arguably most efficient in allowing one to isolate the true propagating
degrees of freedom of general relativity in the full non linear context.
These are in terms of the freely specifiable shear tensors on the
initial null hyper-sufaces, and allows one to uniquely determine the
solution to the Einstein Equations in the vicinity of these null surfaces.
In all that will follow, I shall assume that the spacetime is four
dimensional for definiteness. 

\subsection{Notation and Conventions}

In the following essay, I will use the $(-,+,+,+)$ signature convention.
For concreteness, I implicitly assume there are four spacetime dimensions
although this shall not be a necessary condition for the work that
follows. Greek letters $\mu,\nu,\cdots$ shall run from 0 to 3, capital
Latin indices $A,B,\cdots$ will the values $0$ or $1$ to denote
the two null directions and lowercase Latin indices $a,b,\cdots$
will run from $2,3$ and these are spatial indices. 

I will denote covariant derivatives as $\nabla_{\cdot}$ (the $(\cdot)$
denotes an arbitrary index) , partial derivatives as $\partial_{\cdot}$
and Lie derivatives with respect to an arbitrary vector field $v^{\cdot}$
as $\mathcal{L}_{v^{\cdot}}$.

\subsection{Space-like surfaces in Spacetime}

In this section I will review some rudimentary facts about a compact
two dimensional space like surfaces living in four dimensional spacetime
with metric $g_{\mu\nu}$. As I'll show, this will be useful in understanding
the situation where spacetime is foliated by such surfaces, which
will be the subject of study in the next section. The two dimensional
surface $\mathcal{S}$, in spacetime is `located' at the intersection
of space like and time-like hyper surfaces $\mathcal{M}$, $\mathcal{B}$
respectively, both of which are three dimensional as shown in figure
1.

\begin{figure}
\includegraphics{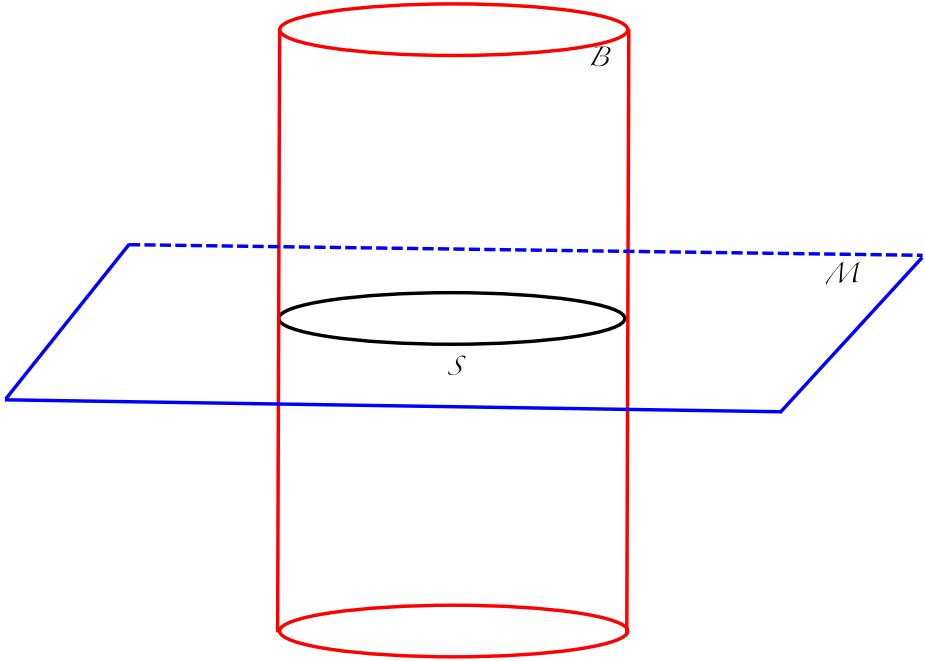} 
\caption{$\mathcal{S}$ is at the intersection of the time-like surface $\mathcal{B}$ and space like surface $\mathcal{M}$ of co-dimension 1}
\label{fig:fig1} 
\end{figure}

This means that the space of normal vectors at a point $p\in\mathcal{S}$,
which is denoted as \footnote{This is a single fiber of the normal bundle $T\mathcal{S}^{\perp}$
to $\mathcal{S}$} $T_{p}\mathcal{S}^{\perp}$ has one space like and one timeline direction.
This means the signature of $T\mathcal{S}^{\perp}$ is $(-,+)$. So
a natural basis of vectors that span this space is the orthonormal basis
of vectors $(m^{\mu},s^{\mu})$ where $m^{\mu}$ is time-like and
$s^{\mu}$ is space like: 
\begin{eqnarray*}
\\
m^{\mu}m_{\mu}=-1,\\
s^{\mu}s_{\mu}=1,\\
m^{\mu}s_{\mu}=0.
\end{eqnarray*}
This choice is by no means unique, because these directions can be
changed by boosts in an arbitrary direction normal to $\mathcal{S}$
which will lead to a new pair of basis vectors $(m'^{\mu},s'^{\mu})$
\[
\left(\begin{array}{c}
m'^{\mu}\\
s'^{\mu}
\end{array}\right)=\left(\begin{array}{cc}
\cosh\alpha & \sinh\alpha\\
\sinh\alpha & \cosh\alpha
\end{array}\right)\left(\begin{array}{c}
m^{\mu}\\
s^{\mu}
\end{array}\right),
\]
and we can check that the conditions $m'^{\mu}m'_{\mu}=-1$, $s'^{\mu}s'_{\mu}=1$,
$m'^{\mu}s'_{\mu}=0$ are still satisfied.

Although by introducing global foliations of spacetime by such time-like
and space-like hyper surfaces, this ambiguity could be fixed. But
the route I will follow is to find what best can be done with whatever
I have introduced above to find a slightly more unique prescription
of a basis on $T_{p}\mathcal{S}^{\perp}$. Recall that the $(-,+)$
signature that $T_{p}\mathcal{S}^{\perp}$ has implies that it possesses
two null directions. This translates to having a null basis formed
by the vectors $(l^{\mu},k^{\mu})$ defined as 
\begin{eqnarray*}
\\
l^{\mu}=\frac{m^{\mu}+s^{\mu}}{\sqrt{2}},\\
k^{\mu}=\frac{m^{\mu}-s^{\mu}}{\sqrt{2}}.
\end{eqnarray*}
It is then easy to see that 
\begin{eqnarray*}
\\
l^{\mu}l_{\mu}=0=k^{\mu}k_{\mu},\\
l^{\mu}k_{\mu}=-1.
\end{eqnarray*}
So it is certainly true that both $l^{\mu}$ and $k^{\mu}$ are null
but they are nowhere parallel to each other. Figure 2 demonstrates
the effect of change in basis. 

\begin{figure}
\includegraphics{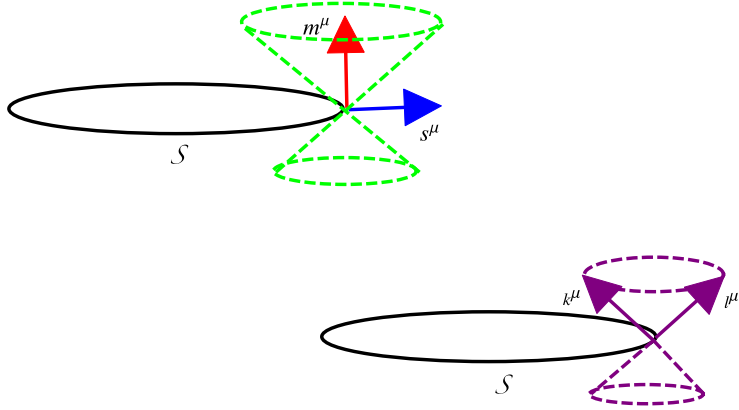} \caption{The space of normal vectors $T_{p}\mathcal{S}^{\perp}$ possesses two null directions}
\end{figure}

Now, the transformation properties for $m^{\mu}$ and $s^{\mu}$ under
boosts translate into the following transformation properties for
the null basis vectors $l^{\mu}$ and $k^{\mu}$: 
\[
\left(\begin{array}{c}
l'^{\mu}\\
k'^{\mu}
\end{array}\right)=\left(\begin{array}{cc}
e^{\alpha} & 0\\
0 & e^{-\alpha}
\end{array}\right)\left(\begin{array}{c}
l^{\mu}\\
k^{\mu}
\end{array}\right),
\]
and so we see that the null vectors only change by a rescaling, and
the null directions themselves remain invariant.

A reasonable question to ask at this point is how the picture where
$\mathcal{S}$ is located at the intersection between the space like
surface $\mathcal{M}$ and the time-like surface $\mathcal{B}$ changes
with this change of basis. The answer is that $\mathcal{S}$ can alternatively
located at the intersection of two null hyper surfaces $\Sigma_{0}$
and $\Sigma_{1}$ to which $l^{\mu}$ and $k^{\mu}$ are tangent,
as is shown in figure 3. 
\begin{figure}
\includegraphics{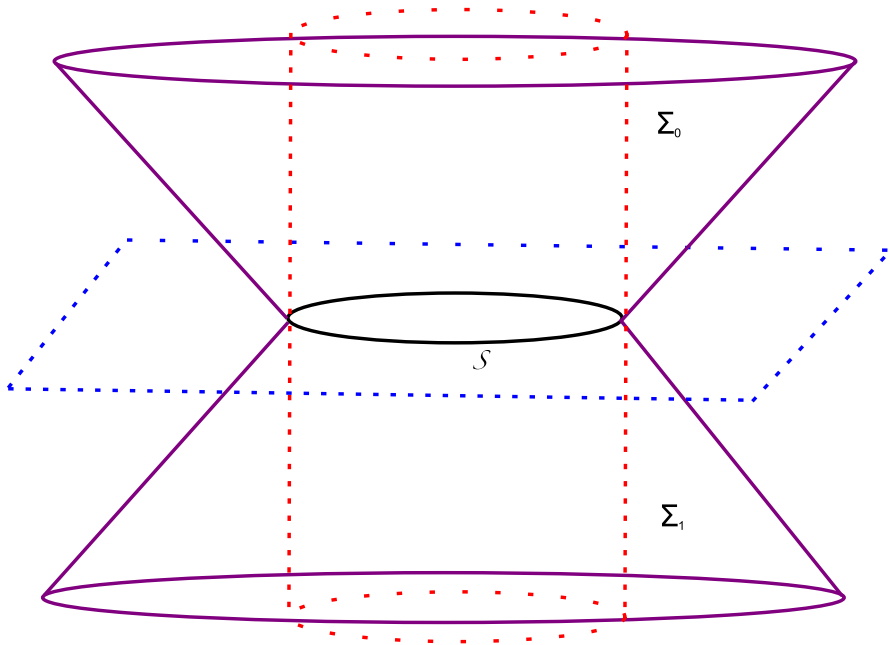} \caption{The null hyper-surfaces $\Sigma_{o}$ and $\Sigma_{1}$ of co-dimension 1 intersect at $\mathcal{S}$}
\end{figure}

Note that if $l^{\mu}$ and $k^{\mu}$ are tangential to $\Sigma_{0}$
and $\Sigma_{1}$ respectively, they are also normal to these hyper-surfaces.
This fact simply follows from the fact that $l^{\mu}l_{\mu}=0=k^{\mu}k_{mu}$
so these vectors are orthogonal to themselves. Intuitively, what this
picture is telling us is that the space-like surface $\mathcal{S}$
is seen to be situated at the intersection of a bundle of light-rays
going into and emanating from it. These bundles of light rays are
null geodesic congruences, which as we shall see in the following
subsection, are the null hyper surfaces $\Sigma_{0}$ and $\Sigma_{1}$.
Now consider spacetime foliated by such space-like two surfaces. Namely,
a two parameter family of such surfaces all located at the intersection
of two null hyper surfaces of dimension three. This is what is known
as the double null foliation.

\subsection{The Double Null Foliation}

I will show how we can upgrade the previous subsection's treatment
of a single space-like surface into a foliation of spacetime by such
intersecting null surfaces. Alternatively, the double null foliation
is mathematically described using a set of embedding relations which
allow us to make the following identification: 
\[
x^{\alpha}=x^{\alpha}(u^{A},y^{a}).
\]
Here $x^{A}=u^{A},A=0,1$ is a scalar function whose level sets are
the null hyper surfaces $\Sigma_{0}^{\tau}$, $\Sigma_{1}^{\tau}$
and $x^{a}=y^{a},a=2,3$ are co-ordinates on the intersection of any
two null surfaces at $\mathcal{S_{\tau}}$. The superscript $\tau$
here is meant to denote that there is a two parameter family of these
null hyper surfaces which form the foliation. Figure 4 aids in visualising
this situation for two leaves of the foliation. 
\begin{figure}
\includegraphics{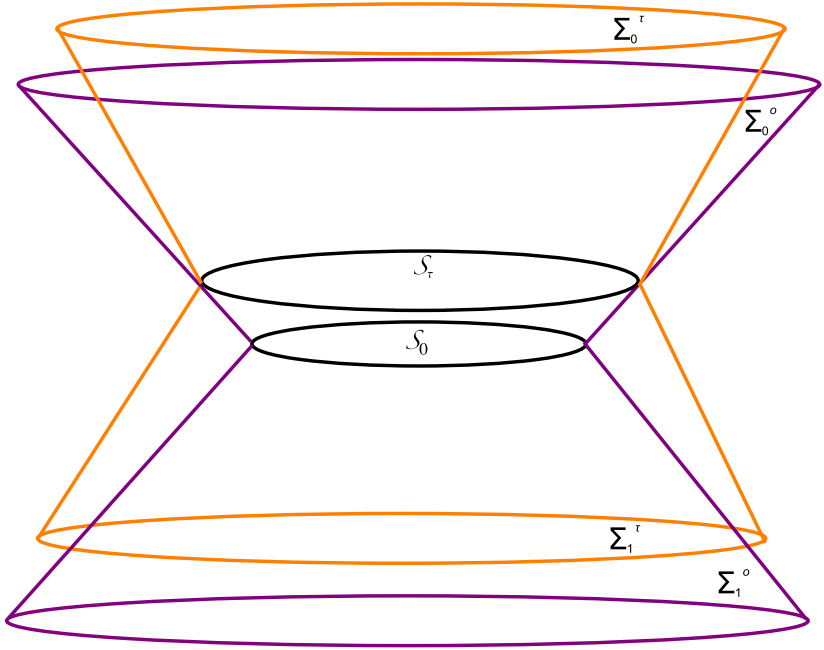}  \caption{Two leaves of the double null foliation of spacetime}
\end{figure}

I will return to the physical interpretation of $\Sigma_{A}$ as initial
value hyper surfaces in the following subsection.

The null character of the $\Sigma_{A}$ hyper surfaces is seen by
noting that: 
\[
\nabla^{\mu}u^{A}\nabla_{\mu}u^{B}=g^{\alpha\beta}\partial_{\alpha}u^{A}\partial_{\beta}u^{B}=e^{-\lambda}\varsigma^{AB}.
\]
Here, $\varsigma^{AB}=\textrm{antidiag}(-1,-1)=\varsigma_{AB}$. We
can use this as a metric to raise and lower the capital latin indices.
What is meant by saying that the null character of the hyper surfaces
is captured by this equation can be seen by identifying the normal
vectors to these hyper surfaces through 
\begin{eqnarray*}
n_{\mu}^{A}=e^{\lambda}\nabla_{\mu}u^{A},\\
n_{A}^{\mu}n_{\mu}^{B}=e^{\lambda}\delta_{B}^{A},\\
n_{A}^{\mu}n_{\mu A}=0\ (\textrm{no sum on }A),
\end{eqnarray*}

In other words, list of conditions may well be summarised as 
\begin{equation}
n^{A}\cdot n^{B}=e^{\lambda}\varsigma^{AB},
\end{equation}
where the dot denotes contraction of the spacetime indices using the
metric $g_{\alpha\beta}$. 
\begin{figure}
\includegraphics{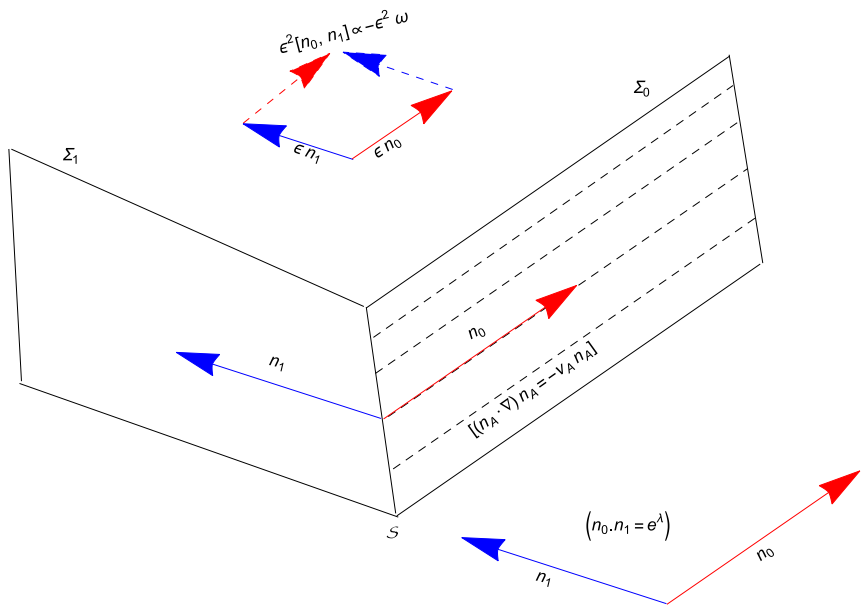}\caption{This diagram summarises the geometrical role of all the quantities
introduced in this subsection. Note that although the space like hyper
surface is compact, we have `unwrapped' it into a line here for purposes
of visualization only.}
\end{figure}

To make the connection with quantities identified in the previous
subsection, we see that per leaf of the foliation: 
\begin{eqnarray*}
\\
l^{\mu}=e^{-\frac{\lambda}{2}}n_{0}^{\mu},\\
k^{\mu}=e^{-\frac{\lambda}{2}}n_{1}^{\mu}.
\end{eqnarray*}
The only reason I introduce the vectors $n_{A}^{\mu}$ in this manner
is to expose the quantity $\lambda$ which, as we will see, is a dynamical
variable. Also $\textrm{exp}(\lambda)$ measures said deviation of
these vectors from being parallel.

A further consequence of these properties is that there exists a complementary
description of this setting in terms of a congruence of null geodesics.
This can be seen by noting that the null vectors $n_{A}^{\mu}$ themselves
satisfy a geodesic equation of the form 
\begin{equation}
\nabla_{n_{A}}n_{A}:=(n_{A}\cdot\nabla)n_{A}=-\nu_{A}n_{A},
\end{equation}
and as is evident from the non zero right hand side, the parameterisation
of these geodesics aren't in general affine. The parameter $\nu_{A}$
is known as the `in-affinity ' and as the name suggests, it measures
the deviation from affinity of the parameterisation of the geodesics.
If we return to the definition of the normal vectors in terms of derivatives
of the optical functions $u^{A},$ we see that an alternative definition
of the in-affinity parameter can be given as $\nu_{A}=\partial_{A}\lambda$.
Exploiting this fact, we can alternatively think of the normals as
the generators of the null hyper surfaces.

Another quantity of interest is the so called normalised twist vector
which is defined as 
\begin{equation}
\omega_{a}=-e^{\lambda}[n_{0},n_{1}]_{a},
\end{equation}
 where $[\cdot,\cdot]$ denotes the Lie brackets. Geometrically, it measures the failure of integrability in the
sense of Frobenius' theorem of the time like hyperplanes orthogonal
to the intersection of the null surfaces. Closely related quantities
which are also convenient to name for future reference are: $\omega_{a\pm}=\pm\omega_{a}+\lambda_{,a}$
and $\zeta_{a\pm}=\pm\omega_{a}+(e^{\lambda})_{,a}$. Briefly shifting
our attention to the intrinsic geometry of the space like surfaces,
we first define the following dyad: 
\[
e_{(a)}^{\alpha}=\frac{\partial x^{\alpha}}{\partial y^{a}},
\]
through which one may define an intrinsic metric on the space like
surface through 
\begin{equation}
e_{(a)\alpha}e_{(b)\beta}g^{\alpha\beta}=\gamma_{ab}.
\end{equation}
The orthogonality of the normal vectors to $\mathcal{S}$ implies
\[
n_{\alpha}^{A}e_{\beta(b)}g^{\alpha\beta}=0.
\]
This, in addition to the fact that $n_{\mu}^{A}=e^{\lambda}\frac{\partial u^{A}}{\partial x^{\mu}}$,
$n^{A}\cdot n_{B}=e^{\lambda}\delta_{B}^{A}$, implies that 
\[
\frac{\partial x^{\mu}}{\partial u^{A}}=n_{A}^{\mu}-s_{A}^{a}e_{(a)}^{\mu}.
\]
The vectors $s_{A}^{a}$ are called the `shift' vectors and the above
condition aptly highlights their geometric role in propagating the
co-ordinatization of the space like `cross sections' of the null hyper
surfaces to which they are tangential everywhere, as we move along
the direction of the generators. Now we can suitably write down the
entire decomposition of the spacetime metric adapted to this foliation
as follows 
\begin{equation}
g_{\mu\nu}(u^{A},y^{a})=\textrm{exp}(-\lambda)\varsigma_{AB}n_{\mu}^{A}n_{\nu}^{B}+\gamma_{ab}e_{\mu}^{(a)}e_{\nu}^{(b)}.
\end{equation}
So, component wise, I summarise the decomposition as follows:

\[
g_{AB}=\textrm{exp}(-\lambda)\varsigma_{AB}+\gamma_{ab}s_{A}^{a}s_{B}^{b}
\]
\[
g_{Ab}=s_{Ab}
\]
\[
g_{ab}=\gamma_{ab}.
\]
To conclude our discussion of the intrinsic geometry of the space
like surfaces, note that we can find the intrinsic spin connection
$\left(\vartheta_{(b)}^{(a)}\right)$ and curvature two form $\left(\mathcal{R}_{(b)}^{(a)}\right)$
using the Cartan structure equations: 
\begin{eqnarray*}
\textrm{d}e^{(a)}+\vartheta_{(b)}^{(a)}\wedge e^{(b)}=0,\\
\textrm{d}\vartheta_{(b)}^{(a)}+\vartheta_{(c)}^{(a)}\wedge\vartheta_{(b)}^{(c)}=\mathcal{R}_{(b)}^{(a)},
\end{eqnarray*}
and so the intrinsic Riemann curvature tensor of $\mathcal{S}$ can
then be found: 
\begin{equation}
e_{(a)}^{d}e_{(b)c}\mathcal{R}_{ef}^{(a)(b)}=\ ^{(2)}R_{cef}^{d},
\end{equation}
from this, we can then deduce the intrinsic Ricci tensor $\ ^{(2)}R_{ef}$
and the scalar curvature $\ ^{(2)}R$.

I will now return to quantities pertinent to the extrinsic geometry
of the space like intersection of the null surfaces embedded in spacetime.
First I introduce some notation. I will denote the Lie derivative
of functions, vectors and tensors living on $\mathcal{S}$ restricted
onto $\mathcal{S}$, i.e. pulled back onto the two surfaces as follows
\[
\partial_{A}=\perp\mathcal{L}_{n_{A}^{a}},
\]
here $\perp$ denotes the operation of projecting everything to its
right onto $\mathcal{S}$ through the dyads $e_{(a)\alpha}$ defined
through $e_{(a)\alpha}e_{(b)\beta}g^{\alpha\beta}=\gamma_{ab}$. The
Extrinsic Curvature of the space like two surfaces is given in terms
of the time derivatives of the metric as one would expect: 
\begin{equation}
K_{Aab}=\frac{1}{2}\perp\mathcal{L}_{n_{A}}\gamma_{ab}=\frac{1}{2}\partial_{A}\gamma_{ab},
\end{equation}
and the shear is defined as the traceless part of the same: 
\[
\sigma_{Aab}=K_{Aab}-\frac{1}{2}K_{A}\gamma_{ab}.
\]
The expansion scalar for the null congruences is given by 
\[
\theta_{A}=\frac{1}{\sqrt{\gamma}}\partial_{A}\sqrt{\gamma}.
\]

The Einstein Hilbert Action reads 
\begin{equation}
S=\int\textrm{d}u^{0}\textrm{d}u^{1}\textrm{d}^{2}x\sqrt{\gamma}\left\{ e^{\lambda}\ ^{(2)}R-\sigma_{ab}^{A}\sigma_{A}^{ab}+\frac{1}{2}\theta^{A}(\theta_{A}-2\nu_{A})-\frac{1}{2}e^{-\lambda}\zeta_{a\pm}\zeta_{\mp}^{a}\right\} .\label{eq:dnac}
\end{equation}
It would also be useful to note that the boundary terms in the Lagrangian
are: 
\begin{equation}
\partial_{\alpha}\left(\sqrt{-g}n_{A}^{\alpha}e^{\lambda}(2\theta^{A}+\nu^{A})\right)+2\partial_{\alpha}\left(\sqrt{-g}e_{(a)}^{\alpha}\lambda^{,a}\right).
\end{equation}
Note that although we try and remain democratic with both null directions
by using the uppercase latin index, if we wish to understand evolution,
I will treat one of the null directions as though it were a space
like direction and the other is considered `time'. The momenta conjugate
to the two metric can be identified as 
\[
\Pi_{A}^{ab}=\frac{\delta L}{\delta(\mathcal{L}_{n_{A}}\gamma_{ab})},
\]
\begin{equation}
\Pi_{A}^{ab}=\frac{1}{2}\sqrt{\gamma}\left(-2\mathcal{L}_{n_{A}}\lambda\right)+\frac{1}{2}\sqrt{\gamma}\left(\gamma_{ab}\gamma_{cd}-\gamma_{ac}\gamma_{bd}\right)\mathcal{L}_{n_{A}}\gamma^{cd},
\end{equation}
and taking the trace of the above we find that 
\[
\textrm{tr}(\Pi_{Aab})=\Pi_{A}=\sqrt{\gamma}(\theta_{A}-2\nu_{A}).
\]
I will repeatedly use this relation in the future. In the forthcoming
section, I will move beyond the kinematical description of this foliation
and attempt to demonstrate its utility in terms of simplifying the
dynamics of general relativity through the solution of the characteristic
initial value problem.

\subsection{Gauge Fixing and Null Evolution}

The following construction will seem slightly more intricate than
the more conventional 3+1 split of spacetime into space like surfaces
but the advantage will be that it is relatively simple to impose gauge
fixing conditions. The evolution problem is rendered free, whereas
data specified on a space like hyper surface needs to satisfy a set
of elliptic constraints that are in general difficult to solve. What
I will sketch in this subsection is what such gauge fixing conditions
are and what initial data to specify and where to do so in order to
uniquely determine the solution to the vacuum Einstein Equations in
a neighbourhood of a pair of intersecting null hyper surfaces. This
was first achieved by Sachs in (\cite{CIVP}). In this subsection however, I shall closely follow the exposition of the solution to the characteristic initial value problem presented in (\cite{BDIM}).

Without loss of generality, I can choose $u^{0}$ to be the `time'
and so $\Sigma_{0}^{o}(u^{0}=0)$ is the initial hyper surface and
$\Sigma_{1}^{o}(u^{1}=0)$ is the `boundary'. Then the normalised
vector fields $L_{0}=l^{\mu}\partial_{\mu}$ and $L_{1}=k^{\mu}\partial_{\mu}$
are such that they satisfy 
\[
L_{0}u^{0}=0=L_{1}u^{1}
\]
\[
L_{1}u^{0}=1=L_{0}u^{1}.
\]
Recall the normalization condition: 
\[
g_{\mu\nu}l^{\mu}k^{\nu}=-1.
\]
The local coordinates on the space like plane $\mathcal{S}$ should
be convected along either null direction, so the gauge fixing conditions
that ensure this are given by setting on $\Sigma_{0}^{o}$ $s_{A}^{a}=0$,
and setting $s_{0}^{a}$ to zero everywhere. Furthermore, we can fix
the normalisation of the vector fields by choosing $\theta_{A}=2\nu_{A}$
or some other condition that fixes $\lambda$.

\begin{figure}
\includegraphics{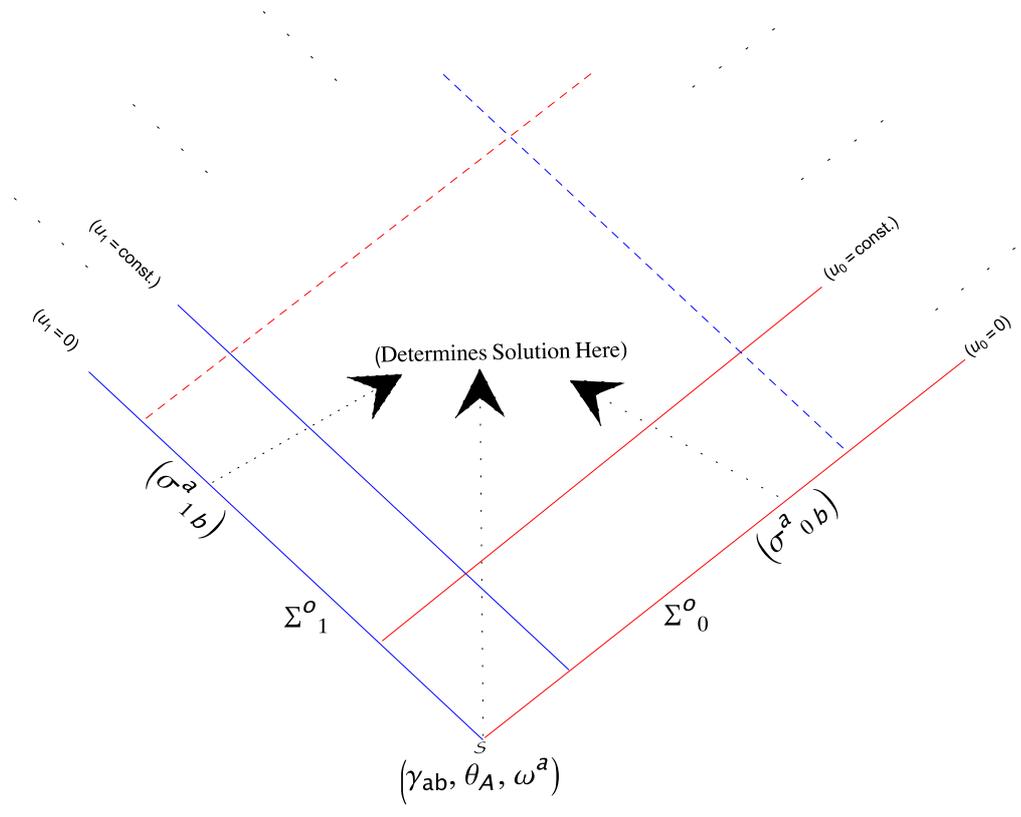}\caption{Summary of where to specify various quantities to determine the solution
to the field equations in the vicinity of $\Sigma_{A}$, at least
locally.}
\end{figure}

With this, I can now write 
\[
\frac{\partial}{\partial u^{0}}=L_{1},
\]
\[
\frac{\partial}{\partial u^{1}}=L_{0}+s_{1}^{a}\partial_{a},
\]
then the conditions $[\partial_{u^{0}},\partial_{u^{1}}]=0=[\partial_{u^{0}},\partial_{a}]$
imply that 
\[
[L_{0},L_{1}]=\frac{\partial s_{1}^{a}}{\partial u^{0}}\partial_{a}.
\]
Thus we see that $\omega^{a}=\partial_{0}s_{1}^{a},$ and although
we have placed a lot of gauge fixing conditions to arrive at this
expression, we note that this expression generalises to an equivalent
definition of $\omega^{a}$ even in the gauge unfixed case, and this
is 
\begin{equation}
\omega^{a}=\epsilon^{AB}(\partial_{B}s_{A}^{a}-s_{B}^{b}s_{A;b}^{a}).
\end{equation}
See (\cite{BDIM}) for a derivation of this relation. 
The gauge fixed, preferred double null foliation is now defined as
the pair of families of intersecting null hyper surfaces given by
\begin{equation}
\Sigma_{0}^{\tau}=\left\{ u^{0}=\tau\right\} ;\Sigma_{1}^{\tau}=\left\{ u^{1}=\tau\right\} .
\end{equation}
Now I will sketch the answer to the second question which is what
data need be specified and where in order to uniquely determine the
solution to the Einstein field equations. First, on $\mathcal{S}$
the functions 
\begin{equation}
(\gamma_{ab},\theta_{A},\omega^{a}),
\end{equation}
need be specified. Then we see that the condition $\theta_{A}=2\nu_{A}$
propagates the choice of $\lambda$ and fixes it along the entire
surfaces $\Sigma_{0}^{o}$ and $\Sigma_{1}^{o}$. We then need to
specify 
\begin{equation}
\sigma_{0b}^{a}\textrm{ on }\Sigma_{0}^{o}\textrm{ },\textrm{ }\sigma_{1b}^{a}\textrm{ on }\Sigma_{1}^{o},
\end{equation}
this is equivalent to specifying the conformal two metric on these
surfaces. These shear rates are the two physical degrees of freedom
of the gravitational field. In all that shall follow I will try and
exploit the fact that the conformal two metric or the conformally
covariant shear rates on null surfaces encode the physical gravitational
degrees of freedom, or in other words, the nonlinear generalisation
of the graviton modes. We then use the dynamical equations $R_{00}=0=R_{0a}$
to propagate $\theta_{0}$ and $\omega^{a}$ along the whole of $\Sigma_{1}^{o}$.
More explicitly, these two equations read (assuming the condition
$\theta_{A}=2\nu_{A}$ has already been imposed) 
\begin{equation}
R_{00}=\partial_{0}\theta_{0}+\theta_{0}^{2}+\frac{1}{2}\sigma^{ab}\sigma_{ab}=0
\end{equation}
\begin{equation}
R_{0a}=-\frac{1}{2}e^{-\lambda}\partial_{0}\omega_{a}-\frac{1}{4}\theta_{0}(\theta_{0}+e^{-\lambda}\omega_{a})-\frac{3}{4}\theta_{0,a}+K_{0a;b}^{b}=0.
\end{equation}
Notice that these are both first order differential equations
for the quantities $\theta_{0}$ and $\omega^{a}$. The two constraint
equations $R_{11}=0=R_{1a}$ similarly propagate these functions onto
the whole initial value surface $\Sigma_{0}^{o}$. Then, the dynamical
equation 
\begin{multline}
R_{ab}=e^{-\lambda}\left\{ 2(\partial_{1}-\mathcal{L}_{s_{1}^{a}})K_{0ab}-K_{1}K_{0ab}\right\} +\frac{1}{2}\ ^{(2)}R\gamma_{ab}+4K_{0(a}^{d}K_{1b)d}\\
+\omega_{(a;b)}-\frac{1}{2}e^{-2\lambda}\omega_{a}\omega_{b}-\lambda_{;ab}-\frac{1}{2}\lambda_{,a}\lambda_{,b}=0
\end{multline}
determines $K_{0ab}$ as a function of $u^{1}$ on $\Sigma_{0}^{\tau}$
for some $\tau$. For data determined by the solutions to the previous
constraint and evolution equations, and the boundary value of $K_{0ab}$
on $\Sigma_{0}^{o}$, these first order ordinary differential equations
can be solved uniquely and in all, the solution to the vacuum field
equations can be uniquely determined within a neighbourhood of these
surfaces as initially advertised.

\subsection{The Conformally Coupled Scalar Field}
In this section I will discuss properties of the theory of a scalar field conformally coupled to General Relativity, this theory is also known as dilaton gravity. To attain its action, I use the fact that the Lagrangian can be obtained form the Einstein--Hilbert
Lagrangian as: 
\[
L[g_{\mu\nu}]\rightarrow\tilde{L}[e^{2\varphi}g_{\mu\nu}],
\]
where $\varphi$ is the scalar field. The resulting action is:
\begin{equation}
S=\int\textrm{d}^{4}x\sqrt{-g}\left(Re^{2\varphi}+6g^{\alpha\beta}\partial_{\alpha}\varphi\partial_{\beta}\varphi\right).\label{eq:cdg}
\end{equation}
Aside from diffeomorphism invariance this action also possesses invariance
under conformal transformations of the form $g_{\alpha\beta}\rightarrow e^{2\psi}g_{\alpha\beta}$
;$\varphi\rightarrow\varphi-\psi$. 
This implies for the two metric: 
\begin{equation}
\gamma_{ij}\rightarrow\tilde{\gamma}_{ab}=e^{2\varphi}\gamma_{ab}.
\end{equation}
So: 
\begin{equation}
\tilde{\Pi}_{A}^{ab}=\frac{\delta\tilde{L}}{\delta(\mathcal{L}_{n_{A}}\tilde{\gamma}_{ab})}=\frac{\delta\tilde{L}}{\delta(\mathcal{L}_{n_{A}}\gamma_{ab})}\frac{\delta(\mathcal{L}_{n_{A}}\gamma_{ab})}{\delta(\mathcal{L}_{n_{A}}\tilde{\gamma}_{ab})}=e^{-2\varphi}\Pi_{A}^{ab}.
\end{equation}
The canonical conjugate of the scalar field can also be identified
straightforwardly as follows: 
\[
\Pi_{\varphi A}=\frac{\delta\tilde{L}}{\delta(\mathcal{L}_{n_{A}}\varphi)}=\frac{\delta\tilde{L}}{\delta(\mathcal{L}_{n_{A}}\tilde{\gamma}_{ab})}\frac{\delta(\mathcal{L}_{n_{A}}\tilde{\gamma}_{ab})}{\delta(\mathcal{L}_{n_{A}}\varphi)}=2e^{\varphi}\gamma_{ab}e^{-2\varphi}\Pi_{A}^{ab},
\]
so we find that, 
\[
\Pi_{\varphi A}=2e^{-\varphi}\Pi_{A}^{ab}\gamma_{ab},
\]
Which can be paraphrased as a constraint 
\begin{equation}
\xi_{A}=2\Pi_{A}^{ab}\gamma_{ab}-e^{\varphi}\Pi_{\varphi A}=0.
\end{equation}
All we need to do now is to simply work out how every object in the
action decomposed in terms of variables adapted to the foliation transforms
under $g_{\alpha\beta}\rightarrow e^{2\varphi}g_{\alpha\beta}$ which
will then be the same as the action for the conformally coupled scalar
decomposed in this manner.

First we will need to understand how the normals transform under this
rescaling. What guides us in the case where the hyper surface is not
null to make this prescription is to demand that the normals still
remain normalised, i.e. $n_{A}^{a}n_{a}^{B}=\pm\delta_{A}^{B}=\tilde{n}_{A}^{a}\tilde{n}_{a}^{B}$,
and knowing how $g_{ab}$ transforms tells us that $\tilde{n}_{Aa}=e^{-\varphi}n_{Aa}$.
Given that now the normals are null, which means that $n_{A}^{a}n_{a}^{B}=0=\tilde{n}_{A}^{a}\tilde{n}_{a}^{A}$,
the vector $n^{a}_{A}$ can scale neutrally, i.e. $\tilde{n}^{a}_{A}=n^{a}_{A}$. However, since
$\lambda$ defined as $g_{ab}n_{0}^{a}n_{1}^{b}=e^{-\lambda}$ is
indeed considered dynamical in this formalism, we can find how it
is to rescale: 
\[
\tilde{g}_{ab}n_{0}^{a}n_{1}^{b}=e^{-\tilde{\lambda}},
\]
and so 
\[
\tilde{\lambda}=\lambda-2\varphi.
\]
In a similar manner we also find that $\tilde{s}_{A}^{a}=s_{A}^{a}.$
Recalling the definition of the normalised twist, it isn't hard to
see that $\tilde{\omega}_{a}=e^{\varphi}\omega_{a}$ and so we also
know how $\tilde{\omega}^{a}_{\pm},\tilde{\zeta}^{a}_{\pm}$ transform. With this we can
write 
\begin{equation}
L_{\textrm{Dilaton}}=\tilde{L}[e^{2\varphi}\gamma_{ab},e^{-2\varphi}\Pi_{A}^{ab},\varphi,\Pi_{\varphi A}(=\textrm{tr}\Pi_{A}e^{-\varphi}),\tilde{\zeta}_{\pm a}].
\end{equation}
So far my discussion has pertained to mathematical properties of the
theory described by the action (\ref{eq:cdg}), but I now wish to
ask the question: under what circumstances can we regain General Relativity
from this theory? The standard answer is fairly simple and it is to
consider GR as a \textit{{phase wherein} } $\varphi=0$. This is
to be thought of as a broken symmetry phase as it isn't preserved
under Weyl transformations mentioned above which are a symmetry of
the Dilaton action. The scalar field seen in this light can be thought
of as an auxiliary `spurion' field that compensates for the metric's
Weyl transformation. And the trick we used to transform the variables
associated to the Einstein Hilbert action to find the ones associated
to the Dilaton action is the Stuckleberg mechanism. Obviously, if
one wishes to return to the original description of the theory undoing
such a transformation, setting $\varphi$ to zero is the easiest means
to do so. But we also see that there is a tension with symmetry in
this procedure which is justified by the fact that General Relativity
is not Weyl invariant.

The question then is, can there be a means to try and retain Weyl
invariance and a notion of equivalence with General Relativity? I.e.
is there a description of the dynamics of the theory which still retains
some subset of the symmetries of the Dilaton which are diffeomorphisms
and local Weyl transformations? As we shall see in the coming sections,
it turns out that it is indeed possible to do this by utilising the
double null foliation for the canonical description of the theory,
and the gauge invariances that will survive turns out to be Diffeomorphisms
of the co-dimension two hyperplanes and Weyl transformations of the
2D metric on said planes. This shall however come at the cost of fixing
a preferred null foliation, or in other words, singling out a specific
manner to parameterise the null geodesics of the congruence of light
rays generating the null hyper-sufaces.

In the following sections of the essay, I will demonstrate how this
is achieved precisely. In order to do so it would be useful for me
to introduce some tools to deal directly with the geometry of this
theory's phase space which, as we shall see is the real arena of its
classical dynamics.

\section{The Geometry of Phase Space}

I would like to introduce some concepts pertaining to the geometry
of phase space such as the symplectic form and constraint surfaces.
As advertised in the previous section of this essay, this will be useful
in understanding the symmetry trading mechanism and subsequent construction
of the theory dual to general relativity in the double null foliation.
This is the objective of the first subsection. In the second subsection,
I will employ this machinery to introduce the symmetry trading algorithm.

\subsection{The Symplectic Form}

Firstly, a convenient piece of mathematical machinery to introduce
is the symplectic form. Consider a field on spacetime, or more precisely in a region $M$ in spacetime, with an action
\[
S[\Phi]=\int_{M}\textrm{d}^{4}xL[\Phi],
\]
where $L[\Phi]$ is the so called Lagrangian density, and here, we
consider it a local functional on the space of solutions to the classical
field equations. The variation of the action will have terms proportional
to the Euler Lagrange equations given as follows 
\[
\delta S=\int_{M}\textrm{d}^{4}x\delta\Phi\left(\frac{\delta L}{\delta\Phi}-\partial_{\mu}\left(\frac{\delta L}{\delta(\partial_{\mu}\Phi)}\right)\right)+\int_{\partial M}\textrm{d}S^{\mu}\left(\frac{\delta L}{\delta(\partial_{\mu}\Phi)}\right)\delta\Phi.
\]
$\textrm{d}S^{\mu}$ is the surface element on the boundary. The first
term vanishes by virtue of the equations of motion being satisfied
and the second boundary term can be identified as the so called `Symplectic
Current' when evaluated on the space of solutions, this defines a
one form on the space of solutions to the field equations, namely.
\[
\Theta=\int_{\Sigma}\textrm{d}S^{\mu}J_{\mu}\delta\Phi=\int_{\Sigma}\textrm{d}S^{\mu}\left(\frac{\delta L}{\delta(\partial_{\mu}\Phi)}\right)\delta\Phi.
\]
The boundary of the four dimensional region where we consider the variation of the action in $M$ shall be denoted $\Sigma$ and it is not necessarily a null surface. 
The functional exterior derivative of the current is the so called
Symplectic form which is a two form on the space of solutions to the
field equations: 
\[
\Omega=\delta\Theta=\int_{\Sigma}\textrm{d}S^{\mu}\delta\left(\frac{\delta L}{\delta(\partial_{\mu}\Phi)}\right)\wedge\delta\Phi.
\]
$\delta\Omega=0$, so it is a closed two form on the space of solutions.
In the previous subsection we put some effort into understanding how
the specification of fields and their derivatives on initial value
hyper surfaces uniquely determine the solution to the field equations
(locally in our case of the double null foliation).

Even more generally, the classical solutions to the field equations
are in one to one correspondence with the values of the field and
its derivatives on some initial value hyper surface. If we were to
choose the hyper surface upon which the symplectic current is defined
to be space like, then the normal component of the Lagrangian's functional
derivative with respect to the derivative of the field, i.e. $n^{\mu}\frac{\delta L}{\delta(\partial_{\mu}\Phi)}=\Pi_{\Phi}$,
is the momentum conjugate to the field. And from what we know of classical
mechanics, the values of momenta and the positions for a system uniquely
determines the solutions to the classical equations of motion. So
in this framework we think of the space of solutions as the so called
covariant phase space (which I will call $\Gamma$), on which the
symplectic form is a natural object to study. This quantity is
called covariant because the symplectic form doesn't depend on
the choice of the hyper surface upon which it is evaluated. So if
we choose the hyper surface $\Sigma'$ different from $\Sigma$ to
try and find the symplectic form as we did above, then although the
form of $\Theta'$ will be different, 
\[
\delta\Theta'=\Omega'=\delta\Theta=\Omega
\]

The symplectic current or potential inherits from the action the freedom
of having added to it a total derivative. This is because the action
is invariant under the addition of a total derivative and so the symplectic
potential inherits this property as well. Another way to see this
is to note that the differential $\delta$ satisifies the property
$\delta^{2}=0$. So If we make the shift $\Theta\rightarrow\Theta+\delta f$
where $f$ is a arbitrary phase space function, then $\Omega$ remains
unaltered. A paradigmatic example of where this property can be used
is to perform canonical transformations on phase space. We will
use this later in the construction of the `Linking Theory' for the
symmetry trading algorithm.

\subsection{The Zoology of Constraints and the Constraint Hypersurface}

In this subsection, I will introduce what are known as first and second
class constraints along with their geometrical interpretation. An
issue of paramount importance is that the aforementioned two form
can in general, and for gauge theories in particular be degenerate.
The degenerate directions are associated to the gauge orbits of the
theory. In a phase space framework, gauge invariance of the theory
manifests itself through the presence of first class constraints,
which are functions $C_{I}(\Phi,\Pi_{\Phi})$ of the fields and their
momenta on the phase space. The constraint equation $C_{I}=0$ defines
a constraint surface on the phase space. In order to understand better
the geometry of such a surface, we need to introduce the notion of
a Hamiltonian vector field corresponding to a phase space function
$F(\Phi,\Pi_{\Phi})$, which is a vector field $X_{F}$ on phase space
such that it satisfies 
\[
\Omega(X_{F})=\delta F.
\]
At this point, it will also be useful to introduce the related concept
of Poisson brackets $\left\{ \cdot,\cdot\right\} $, which mathematically
can be seen by applying the sharp musical isomorphism to the symplectic
form, i.e. $\left\{ F,G\right\} =\Omega(X_{F},X_{G}).$ In this manner,
we can find $X_{C_{I}}$ corresponding to the constraints, and the
first class property is the requirement that 
\[
\Omega(X_{C_{I}},X_{C_{J}})=f_{IJ}^{K}C_{K},
\]
where $f_{IJ}^{K}$ are structure functions. The action of these vector
fields on functions on the phase space is to be interpreted as the
infinitesimal action of the gauge transformations to which they correspond.
The fact that the degenerate directions of the symplectic form correspond
to gauge orbits comes from the trivial observations that on the constraint
surface 
\[
\Omega(X_{C_{I}},X_{C_{J}})=0,\textrm{ }X_{C_{I}},X_{C_{J}}\neq0.
\]
This also tells us that the first class constraints are indeed surface
forming, i.e. Frobenius' theorem ensures that the hyper surface they
are tangent to (the constraint surface $C_{I}=0$) is indeed a sub
manifold of the phase space. Strictly speaking, the true, non degenerate
symplectic form is that which arises from the pullback of $\Omega$,
which we ought to call `presympelctic' on to the quotient of the phase
space by these gauge orbits.

There also exist second class constraints that do not correspond to
any gauge invariance of the theory and they aren't surface forming
either. Geometrically, if we find that a set of constraints $F_{I}$
is second class, it means 
\[
\Omega_{A}(X_{F_{I}},X_{F_{J}})=M_{IJ},
\]
where $M_{IJ}$ is an invertible matrix of phase space
functions, then we dub the system $\left\{ F_{I}\right\} $ a second
class set.

The concept of a constraint surface is also not suited for these constrains
for their action can in general lead to having phase space functions
flow off the physical subspace of the phase space, which is the same
as the constraint surface of the first class set. More precisely,
we find that no combination of the Hamiltonian vector fields $X_{F_{I}}$
can be tangent to the surface defined by $F_{I}=0$. This is because
any vector field $Y$ tangent to this surface would have to satisfy
\[
\Omega(X_{F_{I}},Y)|_{F_{I}}=0.
\]
If $Y=a^{J}X_{F_{J}}$, we then notice that 
\[
\Omega(X_{F_{I}},Y)|_{F_{I}}=a^{J}M_{IJ}
\]
and given that $M_{IJ}$ is invertible, we must have $a^{J}=0$ and
that implies that $Y=0$. This automatically implies that the symplectic
form pulled back to the surface defined by $F_{I}=0$ is certainly
non degenerate.

To be able to explicitly write down a symplectic form, what we need
to do is construct the so called Dirac brackets, and apply onto them
the flat Isomorphism. The Dirac brackets are given by 
\[
\left\{ \cdot,\cdot\right\} ^{*}=\left\{ \cdot,\cdot\right\} -\left\{ \cdot,F_{I}\right\} \left(M^{-1}\right)^{IJ}\left\{ F_{J},\cdot\right\} .
\]
Note that here, repeated capital latin indices imply not only summation
but also potentially integration. Then, for any two vector fields
$X_{A},X_{B}$
\[
\Omega(X_{A},X_{B})|_{F_{I}=0}=\left\{ A,B\right\} ^{*}.
\]
An additional property of first class constraints is that they not
only is their Poisson algebra closed, but their brackets with second class constraints also close, or on the constraint surface:
\[
\left\{ C_{I},F_{J}\right\} =0.
\]
Next, I'll show how this applies to situations we are interested in.

\subsection{The Symplectic Form for Gravity in the Double Null Foliation}

Returning to gravity in the double null foliation, the symplectic
current associated to the action (\ref{eq:dnac}), is given
as a sum of two terms: 
\begin{equation}
\Theta=\int_{\Sigma_{\tau}}J_{0}\textrm{d}u^{1}-J_{1}\textrm{d}u^{0}=\Theta_{0}+\Theta_{1},
\end{equation}
and similarly the presympelctic form is given by the exterior derivative
of the same. More explicitly: 
\begin{equation}
\Theta_{A}=\int_{\mathcal{S}}\textrm{d}^{2}x\sqrt{\gamma}(-\Pi_{Bab}\delta\gamma^{ab}-\omega_{\mp a}\delta s_{B}^{a})+\partial_{B}\int_{\mathcal{S}}\textrm{d}^{2}x\sqrt{\gamma}\delta(\textrm{ln}\sqrt{\gamma}-\lambda),\textrm{ }B\neq A.
\end{equation}
This was derived by Epp in \cite{EPP} . In this light, the Euler
Lagrange equation $R_{Ba}=0$ is actually a constraint that generates
diffeomorphisms tangential to the space like hyperplane in disguise.
To see this, we first write it smeared against a vector field as:
\begin{equation}
\Psi_{Ba}(v^{a}):=\int_{S}\textrm{d}^{2}y\Psi_{Ba}v^{a}=\int_{\mathcal{S}}\textrm{d}^{2}x\left(2\nabla^{b}\Pi_{Bab}-\mathcal{L}_{n_{B}}P_{a}^{\mp}\right)v^{a}.
\end{equation}
Here $P_{a}^{\mp}=\sqrt{\gamma}\omega_{\mp a}=\sqrt{\gamma}(\omega_{a}\pm D_{a}\lambda),$
then we find 
\begin{multline}
X_{\Psi_{Ba}(v^{a})}=\int_{\mathcal{S}}\textrm{d}^{2}x\sqrt{\gamma}\left\{ \mathcal{L}_{v^{a}}\gamma_{ab}\frac{\delta}{\delta\gamma_{ab}}+\mathcal{L}_{v^{a}}\Pi_{B}^{ab}\frac{\delta}{\delta\Pi_{B}^{ab}}+\left[v,n_{B}\right]^{a}\frac{\delta}{\delta s_{B}^{a}}\right\} \\
\\
\Omega_{A}(X_{\Psi_{Ba}(v^{a})},X_{\Psi_{Ba}(v'^{a})})=\delta\Theta_{A}(X_{\Psi_{Ba}(v^{a})},X_{\Psi_{Ba}(v'^{a})})=\Psi_{Ba}([v,v']^{a}).
\end{multline}
And so the first class property is made apparent.

The $R_{BB}$ constraints are of the form 
\begin{equation}
\Psi_{Bo}(f)=\int_{\mathcal{S}}\textrm{d}^{2}x\sqrt{\gamma}f(x)\left(\frac{\partial_{B}^{2}\sqrt{\gamma}}{\sqrt{\gamma}}-\frac{1}{2}\Pi_{B}^{ab}\partial_{B}\gamma_{ab}\right).
\end{equation}
The Hamiltonian vector field $X_{\Psi(f)_{Bo}}$corresponding to this
constraint is defined in the standard manner. It is given by 
\begin{multline}
X_{\Psi_{Bo}(f)}=\int_{\mathcal{S}}\textrm{d}^{2}x\sqrt{\gamma}(\mathcal{L}_{fn_{B}}\gamma_{ab}\frac{\delta}{\delta\gamma_{ab}}+\mathcal{L}_{fn_{B}}\left(\Pi_{Bab}-\frac{1}{2}(\sqrt{\gamma}\gamma_{ab}(\theta_{B}-2\nu_{B}))\right)\frac{\delta}{\delta\Pi_{B}^{ab}}\\
-e^{-\lambda}\mathcal{L}_{n_{B}}\lambda\frac{\delta}{\delta\lambda}).
\end{multline}
We see that its action on the two surface quantities such as $\gamma_{ab}$,
$e^{-\lambda}$ is given by 
\begin{equation}
X_{\Psi_{Bo}(f)}\gamma_{ab}=\perp\mathcal{L}_{fn_{B}}\gamma_{ab}
\end{equation}
\begin{equation}
X_{\Psi_{Bo}(f)}e^{-\lambda}=\perp\mathcal{L}_{fn_{B}}e^{-\lambda},
\end{equation}
Intuitively solving the constraint equation $\Psi_{Bo}=0$allows one to integrate
up quantities defined on a section of a null surface onto the entire
null surface, so this infinitesimal action ought to be expected. For quantities that are intrinsic to the cross section,
the infinitesimal action of this constraint ought to generate normal
deformations along the generators, as it apparently does. Furthermore,
we note that the constraint algebra closes as follows 
\begin{equation}
\Omega_{A}\left(X_{\Psi_{Bo}(f)},X_{\Psi_{Bo}(g)}\right)=\Psi_{Bo}\left(f\mathcal{L}_{n_{A}}g-g\mathcal{L}_{n_{A}}f\right).
\end{equation}
The geometrical role of this constraint is, as mentioned before, to
generate normal deformations of quantities on the space like surface
within the null hyper surface.

The algebra of these constraints and their Poisson brackets were first
computed in (\cite{Tor}) in a Hamiltonian context.

If we take a step back and try to interpret our brief exposition of
the solution to the characteristic initial value problem of general
relativity in the previous subsection interpreting $R_{Ba}=0$ as
a constraint, we see that that the sequence of gauge fixing conditions
we imposed still leaves us with 2 dimensional covariance (as one of
the two constraints would remain unaffected by the gauge fixing condition
$s_{0}^{a}=0$).

The evolution equations, assuming we deem the null direction along increasing $u^{0}$ as time, are $R_{11}=0=R_{1b}$ and $R_{ab}=0$. Their form although, is the same as the constant equations.

So even with all the gauge fixing conditions we impose to turn general
relativity into a theory of null evolution of the physical degrees of
freedom, we are still left with a limited subset of the diffeomorphism
invariance of the full theory.

\subsection{Trading of Gauge Invariances}

We finally have the tools to now understand the symmetry trading algorithm.
I will closely follow the treatment in section 3 of (\cite{GGK})
where this technique was first developed. First, I begin by extending
the phase space $\Gamma$ to a larger one $\mathcal{P}$ which is
the augmentation of the original phase space by auxiliary fields and
their momenta. An example of such a field is the scalar field conformally
coupled to gravity, which was discussed towards the end of the first
section of this essay.

From our discussion of constraints in the previous subsection, it
should be clear that in a theory that possesses gauge invariances,
it is the constraint surface which is of physical importance. This
is because it is upon the constraint surfaces that the admissible
solutions to the classical equations of motion live, because by definition
it is where the constraints are satisfied.

I will denote the constraint surface $\Gamma_{1}:=\left\{ x\in\mathcal{P}|F_{I}(x)=0\ \forall I\right\} $.
The reason for the subscript 1 will become apparent soon, and here,
$\left\{ F_{I}\right\} $ denotes a set of first class constraints.
The Hamiltonian vector fields $X_{F_{I}}$ generate the infinitesimal
action of the gauge group $\mathbb{G}$ \footnote{For the precise definition of the notion of group action on symplectic manifolds and such, see for instance \emph{V. I. Arnold, Mathematical Methods of Classical Mechanics, Springer-Verlag (1989), ISBN 0-387-96890-3.}} on $\Gamma_{1}$. The constraint
surface is then isomorphic to a bundle $E$ over $\Gamma_{1}/\mathbb{G}$
with isomorphism $i:E\rightarrow\Gamma_{1}$. The fibres of this bundle
are the gauge orbits of a point $x\in\Gamma_{1}$ under the action
of $\mathbb{G}$, so they are isomorphic to $\mathbb{G}/\textrm{Iso}(x)$,
where $\textrm{Iso}(x)$ denotes the isotropy group of the point $x$.
In this geometric setup, a section $\mathbf{s}$ of the bundle $E$
is the same as a gauge fixing of the first class constraints $i(\mathbf{s})$.

\begin{figure}
\includegraphics[height=75mm]{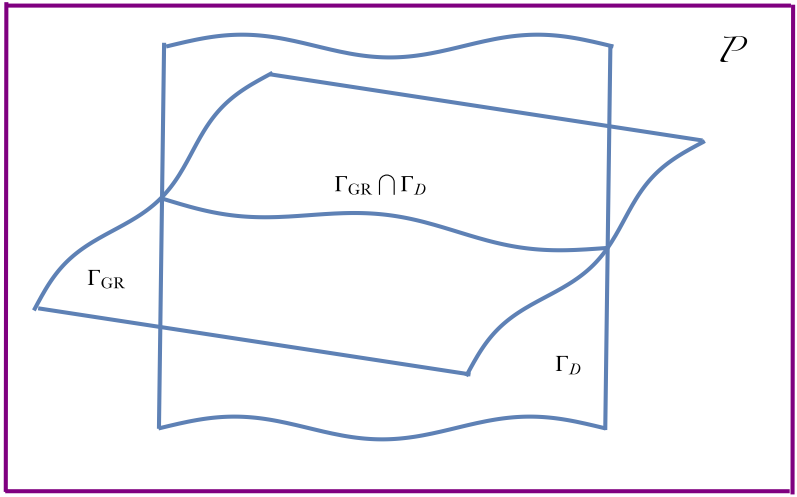} \caption{Intersecting constraint surfaces of the original and dual theories}
\end{figure}

Gauge fixing conditions are a set of functions $\left\{ G_{I}\right\} $
on $\mathcal{P}$ s uch that the intersection of $\Gamma_{2}:=\left\{ x\in\mathcal{P}|G_{I}(x)=0\ \forall I\right\} $
with $\Gamma_{1}$ coincides with $i(\mathbf{s})$. If we then demand
that there exist structure functions $g_{JK}^{I}$ such that $\Omega(X_{G_{J}},X_{G_{K}})=g_{JK}^{I}G_{I}$,
then the dual theory is defined as one which possesses $\left\{ G_{I}\right\} $
as a first class constraint set. This means that we interpret $X_{G_{I}}$
as the generators of the infinitesimal action of some other gauge
group $\mathbb{H}$ on $\Gamma_{2}$. As we saw with $\Gamma_{1}$,
we can then define a bundle $B$ over $\Gamma_{2}/\mathbb{H}$ whose
fibres are $\mathbb{H}/\textrm{Iso}(x)$. This bundle is isomorphic
to $\Gamma_{2}$ with isomorphism $j:B\rightarrow\Gamma_{2}$. 
However, the sets $\left\{F_{I}\right\}$ and $\left\{G_{I}\right\}$ needn't necessarily be disjoint, but it is important that they aren't entirely identical.
Figure 7 illustrates the situation at hand.

I would like to now make some comments on this construction. Note
that because we are using covariant phase space and we interpret the
constraint surface as the space of admissible solutions to the classical
equations of motion, we already see that the gauge fixing $i(\mathbf{s})$
is implicitly preserved under the equations of motion. Moreover, note
that $i(\mathbf{s})$ lies entirely within $\Gamma$, prior to extension
to $\mathcal{P}$, because it is a gauge orbit within the space of
admissible solutions to the equations of motion of the original theory.
The reason why the dual theory can be considered equivalent to the
original theory is because it too contains $i(\mathbf{s})$ within
its constraint surface. This is another way of saying that both theories
share a gauge fixing wherein the initial value problem and equations
of motion are identical. But we also see that in order to construct
a dual theory it was necessary to be able to augment $\Gamma$ by
the phase space of auxiliary fields to obtain $\mathcal{P}$. Although
the common gauge fixing of the two theories, $i(\mathbf{s})$, and
the constraint surface of the original theory $\Gamma_{1}$ lie entirely
within $\Gamma$, the constraint surface $\Gamma_{2}$ of the dual
theory is not restricted to lie in $\Gamma$, and in general it doesn't.
This is why the dual theory is more than just a gauge fixing condition
of the original theory.

The reason this procedure is named symmetry trading is because although
the original and dual theories both share a gauge, the intersection
of their constraint surfaces can be though of as the gauge fixing
of different kinds of gauge invariances in either theory. This is
a consequence of the fact that the gauge groups $\mathbb{G}$ and
$\mathbb{H}$ are distinct, as are the constraint sets $\left\{ F_{I}\right\} $
and $\left\{ G_{I}\right\} $ \footnote{Here, by `distinct', I mean they aren't entirely identical. But as mentioned in an earlier paragraph, they needn't necessarily be disjoint sets}. So we can expect that upon application
of this procedure to the situation we are interested in, which is
that of general relativity in the double null foliation, we should
expect to find a dual theory which possesses gauge symmetries that
are not the same as those of general relativity, or more precisely,
the residual gauge symmetries of the partially gauge fixed theory
describing double null evolution.

We shall see in the coming section that the gauge invariance of the
null evolution problem that is foliation preserving diffeomorphism
invariance will be enhanced by foliation preserving Weyl transformations.
The dual theory will be interpreted as a class of field configurations
of a scalar field conformally coupled to gravity.

\section{The Scale Invariant Theory Dual to Gravity in the Double Null Foliation}

In this section, I will present the trading mechanism applied to the
case of general relativity in the double null foliation to obtain
a dual theory, which, as we will see possesses Weyl and Diffeomorphism
invariance which preserves the foliation. We will also see what the
gauge fixing that needs to be applied here is for the trading to work,
and what the geometrical interpretation of the preferred gauge choice
is.

\subsection{Phase Space Extension and the Dual Theory}

The very first step we need to take before we can begin to apply the
machinery developed in subsection 2.4 is to first extend the phase
space of the theory by auxiliary fields $\varphi$ and their momenta
$\Pi_{\varphi A}$. And, as I mentioned before, this augmentation
in the case we are interested in is by the phase space of the conformally
coupled scalar field. First recall the observation made in subsection
1.5 regarding the field redefinitions to the gravity Lagrangian one
can make in order to attain the theory of a conformally coupled scalar
field. To implement this redefiniton in phase space, we make the simple
observation that if we were to consider a local change of co-ordinates
on the phase space. In those local coordinates, if we write the symplectic
potential as $\hat{\Theta}$ then 
\[
\hat{\Theta}-\Theta=\delta\mathcal{F},
\]
then since $\delta^{2}=0$, 
\[
\hat{\Omega}=\delta\hat{\Theta}=\delta\Theta=\Omega.
\]
Thus this is nothing more than a canonical transformation. Now we
wish to find explicitly a canonical transformation on the extended
phase space of $(\phi,\Pi_{\varphi};\gamma_{ab},\Pi_{A}^{ab})$, where
$e^{\varphi}=\phi$. It turns out that the variable $\xi_{A}=\Pi^{ab}_{A}\gamma_{ab}-e^{\varphi}\Pi_{\varphi A}$ generates
a type two canonical transformation, and this is readily seen by first
considering it smeared against a scalar function $\omega$: 
\[
\xi_{A}(\omega)=\int_{S}\textrm{d}^{2}y\omega\xi_{A}.
\]
Then, 
\begin{multline}
\hat{\Theta}_{A}-\Theta_{A}=\int_{\mathcal{S}}\textrm{d}^{2}x\sqrt{\tilde{\gamma}}\left(-\tilde{\Pi}_{Bab}\delta\tilde{\gamma}^{ab}-\Pi_{\varphi A}\delta\varphi-\tilde{\omega}_{\mp a}\delta s_{B}^{a}\right)+\partial_{B}\int_{\mathcal{S}}\textrm{d}^{2}x\sqrt{\tilde{\gamma}}\delta\left(\textrm{ln}\sqrt{\tilde{\gamma}}+\tilde{\lambda}\right)-\\
\\
\int_{\mathcal{S}}\textrm{d}^{2}x\sqrt{\gamma}(-\Pi_{Bab}\delta\gamma^{ab}-\omega_{\mp a}\delta s_{B}^{a})+\partial_{B}\int_{\mathcal{S}}\textrm{d}^{2}x\sqrt{\gamma}\delta(\textrm{ln}\sqrt{\gamma}-\lambda)=\int_{\mathcal{S}}\textrm{d}^{2}x\sqrt{\gamma}\xi_{A}e^{-\varphi}\delta\varphi
\end{multline}
Where the variables that $\hat{\Theta}_{A}$ consists of are defined
in the previous section. We define the Hamiltonian vector field $X_{\xi_{A}(\omega)}$
associated to $\xi_{A}(\omega)$ through $\Omega(X_{\xi_{A}(\omega)})=\delta\xi_{A}(\omega)$.
Given that this is a generating function, the correposnding Hamiltonian
vector field generates the infinitesimal version of this canonical
trasnformation when applied to the various phase space variables as
follows: 
\begin{multline*}
\\
X_{\xi(\omega)_{A}}\phi=-\omega\phi\\
X_{\xi(\omega)_{A}}\Pi_{\varphi A}=\omega\Pi_{\varphi A}\\
X_{\xi(\omega)_{A}}\gamma_{ab}=2\omega\gamma_{ab}\\
X_{\xi(\omega)_{A}}\Pi_{A}^{ab}=-2\omega\Pi_{A}^{ab},\\
\end{multline*}
here, $\phi=e^{\varphi}$. The action of the type 2 Generating Function
$\xi_{A}$ on the Phase Space variables $(\phi,\Pi_{\varphi A};\gamma_{ab},\Pi_{A}^{ab})$
is thus: 
\begin{equation}
(\phi,\Pi_{\varphi A};\gamma_{ab},\Pi_{A}^{ab})\rightarrow(\phi,\Pi_{\varphi A}-2e^{\varphi}\Pi_{A}^{ab}\gamma_{ab};e^{2\varphi}\gamma_{ab},e^{-2\varphi}\Pi_{A}^{ab}).
\end{equation}
I will call this theory which lives in the extended phase space the
Linking Theory. Why this is a useful observation is that as advertised
before, we now wish to find some analogue of the $\varphi=0$ `phase'
of the theory but one which possesses some subset of the Weyl invariance
of the Dilaton action as well. Such a phase will be given by: 
\begin{equation}
\Pi_{\varphi A}=0
\end{equation}
which is also a gauge fixing of the second class constraint above
that leads to the first class constraint 
\begin{equation}
\mathcal{D}_{A}(\kappa)=2\int_{S}\textrm{d}^{2}y\kappa\Pi_{A}^{ab}\gamma_{ab}.
\end{equation}
The infinitesimal action of this constraint $X_{\mathcal{D}}(\cdot)$
on the phase space variables is that of infinitesimal Weyl transformations:
\begin{equation}
X_{\mathcal{D}_{A}(\kappa)}\phi=-\kappa \phi
\end{equation}
\begin{equation}
X_{\mathcal{D}_{A}(\kappa)}\gamma_{ab}=2\kappa\gamma_{ab}
\end{equation}
\begin{equation}
X_{\mathcal{D}_{A}(\kappa)}\Pi_{A}^{ab}=-2\kappa\Pi_{A}^{ab}.
\end{equation}
So we see that the Weyl transformations on the space-like hyperplane
are preserved in this gauge. The constraint surface for the dual theory
is defined by the functions $\left\{ \mathcal{D}(\kappa)\right\} $
satisfying the algebra: 
\begin{equation}
\Omega(X_{\mathcal{D}(\kappa)},X_{\mathcal{D}(\kappa')})=0.
\end{equation}
What is now left to do is to derive consistency conditions which the
trading algorithm demands certain fields to satisfy. As we will see
shortly, these are in the form of differential equations $\phi$ and
the smearing function for the $\Psi_{Bo}$ constraint (which I previously
dented as $f$) need to satisfy. But before doing this, it will also
be useful to understand the geometrical meaning of the gauge fixing
condition we have imposed above, i.e. the condition that $\textrm{tr}\Pi_{A}=0$.
This will be the purpose of the following subsection.

\subsection{The Gauge Fixing and Second Class Constraints}

Now I will present the geometrical meaning of the gauge fixing condition
imposed, and further consequences of imposing said condition. We see
that the condition 
\begin{equation}
\textrm{tr}\Pi_{A}=0=\theta_{A}-2\nu_{A},
\end{equation}
restricts $\lambda$ and hence the parameterisation of the null geodesics
that generate the null hyper surfaces. Recall the geometric interpretation
associated with the $R_{BB}=0$ constraints, which was that they are
responsible for generating deformations of the space like two surface
along the generators of the generators of the hyper surface. This
can be alternatively interpreted as a statement about the arbitrariness
of the ruling of the null surface. To see why this is the case, imagine
the cross section of the null hyper surface which is specified by
specifying the value of one of the parameters $u^{B}=c$ where $c$
is some constant. If the normal deformation of this two-surface in
the direction of $u^{B}$, or in other words, in parallel to $n_{B}^{a}$,
is to be pure gauge, it means that there exists a one parameter family
of choices of $u'^{B}\neq u^{B}$ for which $u'^{B}=c+\epsilon$ defines
the deformed two surface. The reason for the additional $\epsilon$
in the previous equation is that this deformed two surface is to be
though of in some sense as being `ahead' of the one prior to deformation
along the null direction of the deformation. The one parameter ambiguity
of which I speak hides in the arbitrariness of the function $f(x)$
which we smear against the $\Psi_{Bo}$ constraint.

Before deriving such a condition, it would be useful to note that
because a restriction of this kind exists, the constraint $\Psi_{Bo}(f)$
is no longer first class. In order to proceed I would like to make
some field redefinitions which will simplify forthcoming computations.
We identify the trace free part of the momentum, and find that it
is just the shear tensor: 
\[
\Pi_{A}^{ab}-\frac{1}{2}\gamma^{ab}\textrm{tr}\Pi_{A}=\sigma_{A}^{ab},
\]
in terms of which the $\Psi_{Bo}$ constraint takes the form 
\[
\Psi_{Bo}(f)=\int_{\mathcal{S}}\textrm{d}^{2}x\sqrt{\gamma}f(x)\left(\partial_{B}\theta_{B}+\frac{1}{2}\theta_{B}(\theta_{B}+2\nu_{B})+\frac{1}{2}\sigma_{B}^{ab}\sigma_{abB}\right)
\]
It is now apparent that the constraint equation $\Psi_{Bo}=0$ is the same as the Raychadhuri equation for each of the hyper surfaces thought of as congruences of null geodesics. 
Upon applying the canonical transformation, we find 
\[
\tilde{\sigma}_{A}^{ab}=e^{-2\varphi}\sigma_{A}^{ab}.
\]
Then we can isolate what is called the conformal two metric on $\mathcal{S}$,
defined as: 
\begin{equation}
\gamma_{ab}=e^{2\varphi+\textrm{ln}\sqrt{\gamma}}\frac{h_{ab}}{\sqrt{\gamma_{o}}}=e^{2\rho}h_{ab}.
\end{equation}
here $\rho=\varphi+\frac{1}{2}(\textrm{ln}\sqrt{\gamma}-\textrm{ln}\sqrt{\gamma_{o}})$,
and $\gamma_{oab}$ is a fixed reference metric, whose determinant
is 1. The reference structures are needed to make sure that the conformal
two metric is not a density and will have unit determinant. We denote
$e^{\rho}$ at times as $\varrho$ and $e^{2\rho}$ as $\eta$. The
shear is given very directly in terms of the derivative of the conformal
two metric: 
\[
\partial_{A}h_{ab}=\tilde{\sigma}_{Aab}.
\]
Now we see that the transformed expansion scalar is: 
\[
\tilde{\theta}_{A}=e^{-2\rho}\partial_{A}e^{2\rho}=\frac{\partial_{A}\eta}{\eta}.
\]
Next, I simplify the expression for the canonically transformed constraint
$\tilde{\Psi}_{Bo}$, this is done by first noticing that 
\begin{multline*}
\\
\textrm{tr}\tilde{\Pi}_{A}=\textrm{tr}\Pi_{A}-\Pi_{\varphi}\phi=\xi_{A};\\
\xi_{A}=0=\textrm{tr}\tilde{\Pi}_{A}=>\textrm{tr}\Pi_{A}=\Pi_{\varphi A}\phi\\
=>\tilde{\theta}_{A}-2\tilde{\nu}_{A}=0,\\
\end{multline*}
We can use this to rewrite the constraint as: 
\begin{multline*}
\tilde{\Psi}_{Bo}(f)=\int_{\mathcal{S}}\textrm{d}^{2}x\sqrt{h}f(x)\left(\partial_{B}\tilde{\theta}_{B}+\frac{1}{2}\tilde{\theta}_{B}(\tilde{\theta}_{B}+2\tilde{\nu}_{A})+\frac{1}{2}\sigma_{Bab}\sigma_{B}^{ab}\right)=\\
\int_{\mathcal{S}}\textrm{d}^{2}x\sqrt{h}f(x)\left(\partial_{B}\tilde{\theta}_{B}+\tilde{\theta}_{B}^{2}+\frac{1}{2}\sigma_{Bab}\sigma_{B}^{ab}\right).
\end{multline*}
The conformal invariance of the square of the shear implies that 
\[
\tilde{\sigma}_{Bab}\tilde{\sigma}_{B}^{ab}=\sigma_{Bab}\sigma_{B}^{ab},
\]
which is why I don't put tildes over these variables in the transformed
constraint. Utilising the new definition of the transformed expansion
scalar, I can write this constraint finally in the form 
\begin{equation}
\tilde{\Psi}_{Bo}(f)=\int_{\mathcal{S}}\textrm{d}^{2}xf(x)\left(\partial_{B}^{2}\eta+\frac{\eta}{2}\sigma_{Bab}\sigma_{B}^{ab}\right).
\end{equation}
We see that $\tilde{\Psi}_{Bo}=0$ implies that the conformal factor
has to satisfy the equation 
\begin{equation}
\frac{\partial_{B}^{2}\eta}{\eta}=-\frac{1}{2}\sigma_{Bab}\sigma_{B}^{ab}.\label{eq:lye}
\end{equation}

Now my claim that this constraint would be second class with respect
to $\Pi_{\phi A}=0$ can now be easily checked. First, I write 
\[
\Omega(X_{\Pi_{\varphi A}},X_{\tilde{\Psi}_{Bo}(f)})=\iota_{X_{\Pi_{\varphi A}}}\delta\tilde{\Psi}_{Bo}(f)=X_{\Pi_{\varphi A}}\left(\tilde{\Psi}_{Bo}(f)\right)=\left\{ \Pi_{\varphi A},\tilde{\Psi}_{Bo}(f)\right\} .
\]
Then, by definition 
\[
X_{\Pi_{\varphi A}}=\frac{\delta}{\delta\phi}=\sqrt{\frac{\gamma}{\gamma_{o}}}\frac{\delta}{\delta\varrho},
\]
so I can apply this to compute 
\[
X_{\Pi_{\varphi A}}\left(\tilde{\Psi}_{Bo}(f)\right)=\sqrt{\frac{\gamma(y)}{\gamma_{o}}}\frac{\delta\tilde{\Psi}_{Bo}(f(x))}{\delta\varrho(y)}=\int_{S}\textrm{d}^{2}x\sqrt{\frac{\gamma}{\gamma_{o}}}2\left(\partial_{B}^{2}(\varrho f)+\frac{\varrho f}{2}\sigma_{Bab}\sigma_{B}^{ab}\right).
\]
As promised, the right hand side is neither zero, nor is it proportional
to a constraint. Thus we find that consistency demands that the smearing
function $f(x)$ for the constraint $\tilde{\Psi}_{Bo}$ has to satisfy
the equation 
\begin{equation}
\left(\partial_{B}^{2}+\frac{1}{2}\sigma_{Bab}\sigma_{B}^{ab}\right)\varrho f=0.\label{eq:lfe}
\end{equation}
This implies that this function is no longer a Lagrange multiplier,
but is now fixed by this equation which too is but a manifestation
of the fact that we now have a preferred ruling of the null hyper
surfaces.

Finally, the constraint generating the diffeomorphisms tangential
to the intersection is given by the Euler Lagrange equation $R_{Ba}=0$,
as mentioned and identified before, but we now reproduce it re written
in terms of the field redefinitions and after performing the canonical
transformation: 
\begin{equation}
\tilde{\Psi}_{Ba}(v^{a})=\int_{S}\textrm{d}^{2}x\left(2e^{2\rho}\nabla^{b}\sigma_{Bab}-\mathcal{L}_{n_{B}}\tilde{P}_{a}^{\mp}\right)v^{a}.
\end{equation}
A simplification that has been suppressed is the vanishing of part
of the above constraint given by $(\phi\Pi_{\varphi A})_{,a}=0$ due
to the second class constraint. Also, $\tilde{P}_{a}^{\mp}=e^{\rho}P_{a}^{\mp}.$
The action of the same on the phase space variables is given by: 
\begin{equation}
\left\{ h_{ab},\tilde{\Psi}_{Ba}(v^{a})\right\} =\mathcal{L}_{v^{a}}h_{ab}
\end{equation}
\begin{equation}
\left\{ \tilde{\sigma}_{A}^{ab},\Psi_{Ba}(v^{a})\right\} =\mathcal{L}_{v^{a}}\tilde{\sigma}_{A}^{ab}.
\end{equation}
The algebra of these constraints survives the canonical transformation
as it should, and moreover its brackets with the Weyl constraint
are given by: 
\begin{equation}
\left\{ \tilde{\Psi}_{Aa}(v^{a}),\mathcal{D}_{A}(\kappa)\right\} =\mathcal{D}_{A}(\mathcal{L}_{v^{a}}\kappa).
\end{equation}

What is left to do now is to formally write down the Dirac brackets
for this theory.

\subsection{Dirac Brackets and the Symplectic Form for the Dual Theory}

I will now formally construct the analogue of the matrix $M_{IJ}$
defined in subsection of the previous section for the case of the
dual theory described above. Following appendix A1 of (\cite{Birk})
we note that for a block diagonal matrix of operators of the form
\[
D_{IJ}=\left(\begin{array}{cc}
A & \Delta\\
-\Delta & 0
\end{array}\right),
\]
the inverse is given by 
\[
\left(D^{-1}\right)^{IJ}=\left(\begin{array}{cc}
0 & -\Delta^{-1}\\
\Delta^{-1} & \Delta^{-1}A\Delta^{-1}
\end{array}\right).
\]

We can now write the operator $M_{IJ}$ for the dual theory as 
\begin{equation}
M_{IJ}=\left(\begin{array}{cc}
A(x,y)=\{\tilde{\Psi}_{Bo}(x),\tilde{\Psi}_{Bo}(y)\} & \Delta(x,y)=\{\tilde{\Psi}_{Bo}(x),\Pi_{\varphi B}(y)\}\\
-\Delta(x,y) & 0=\{\Pi_{\varphi B}(x),\Pi_{\varphi B}(y)\}
\end{array}\right).
\end{equation}
Upon formally inverting as was done with the matrix $D$, we find
that the formal Dirac brackets explicitly take the following form:
\begin{multline}
\{\cdot,\cdot\}^{*}=\{\cdot,\cdot\}-\{\cdot,\Pi_{\varphi B}\}\Delta^{-1}\{\tilde{\Psi}_{Bo},\cdot\}-\{\cdot,\tilde{\Psi}_{Bo}\}\Delta^{-1}\{\Pi_{\varphi B},\cdot\}\\
-\{\cdot,\Pi_{\varphi B}\}(\Delta^{-1}A\Delta^{-1})\{\Pi_{\varphi B},\cdot\}
\end{multline}

The symbol $\Delta^{-1}$ denotes the operator inverse of the bracket
between $\Pi_{\varphi B}$ and the constraint $\tilde{\Psi}_{Bo}$,
and the invertibility of the bracket relies on the existence of solution
to the linear differential equation (\ref{eq:lfe}). More precisely,
the operator $\Delta(x,y)$ possesses a non trivial one dimensional
kernel generated by the function $f_{o}$, where the subscript $o$
is meant to denote that it solves (\ref{eq:lfe}). This means that
the following condition is satisfied by the inverse operator 
\[
\int\textrm{d}^{2}x'\Delta^{-1}(x',x)\Delta(x,y)=\delta(x,y)-bf_{o}(x)\sqrt{\gamma}(y),
\]
where $b$ is an arbitrary constant.

Note that the Dirac brackets are equivalent correspond to contracting
the pre symplectic two form with Hamiltonian vector fields of phase
space functions whose dependence on $\varrho$ and $f$ are restricted
to solutions to equations (\ref{eq:lye}), (\ref{eq:lfe}) and we
denote the solutions of said equations $\varrho_{o}$, $f_{o}$ respectively.
Stated mathematically, this implies that 
\begin{equation}
\left\{ \cdot,\cdot\right\} ^{*}=\left\{ \cdot|_{\varrho=\varrho_{o},f=f_{o}},\cdot|_{\varrho=\varrho_{o},f=f_{o}}\right\} .
\end{equation}
This is but a consequence of the fact that 
\[
\left\{ \cdot,\cdot\right\} ^{*}|_{F_{I}=0}=\left\{ \cdot|_{F_{I}=0},\cdot|_{F_{I}=0}\right\} ,
\]
for an arbitrary second class constraint set $\left\{ F_{I}\right\} $.

There is thus a reduction of the Dirac brackets to standard Poisson
brackets when we restrict our attention to this class of phase space
functions whose dependence on $\varrho$ and $f$ are restricted as
mentioned above. On this reduced phase space, we can use the flat
isomorphism on such Hamiltonian vector fields to attain the (pre)symplectic
form on such a reduction. So I take a arbitrary phase space function(al)s
$g,k$ which have some dependence on the functions $\varrho$, $f$,
i.e. $g=g[\varrho,f,x),k=k[\varrho,f,x)$. Then my previous statement
translates to: 
\begin{equation}
\Omega\left(X_{g}|_{f=f_{o},\varrho=\varrho_{o}},X_{k}|_{f=f_{o},\varrho=\varrho_{o}}\right)=\tilde{\Omega}_{A}(X_{g},X_{k}).
\end{equation}
The two form $\tilde{\Omega}_{A}$ is defined to be the symplectic
form for the dual theory.

To summarise, given that we can construct Dirac brackets as I have
shown above, they define a restricted sub space within phase space.
Then consider the Dirac brackets between an arbitrary pair of functions
on phase space, and this becomes the Poisson bracket between the two
functions restricted to the sub space where the second class constraints
are satisfied. The inverse of these Poisson brackets is the definition
of the symplectic form of the dual theory.

Written in local co-ordinates, the symplectic form for the dual theory
is then given by 
\begin{equation}
\tilde{\Omega}_{A}=\int_{\mathcal{S}}\textrm{d}^{2}x\ \delta\tilde{\sigma}_{Bab}\wedge\delta h^{ab}+\delta\tilde{P}_{a}^{\mp}\wedge\delta s_{B}^{a}+\partial_{B}\int_{\mathcal{S}}\textrm{d}^{2}x\ \delta\varrho\wedge\delta\lambda.\label{eq:symp}
\end{equation}
From our discussion in section 2, we know that this thoroughly encapsulates
the classical dynamics of the theory. This concludes the construction
of the dual theory. 

To summarise, the dual theory possesses the symplectic form (\ref{eq:symp}) and the dynamics is dictated by the action 
\begin{equation}
S=\int\textrm{d}u^{0}\textrm{d}u^{1}\textrm{d}^{2}x\sqrt{h}\left\{ e^{\tilde{\lambda}}(\ ^{(2)}\tilde{R}\phi+h^{ab}\partial_{a}\phi\partial_{b}\phi)-\sigma_{ab}^{A}\sigma_{A}^{ab}+\frac{1}{2}\tilde{\theta}^{A}(\tilde{\theta}_{A}-2\tilde{\nu}_{A})-\frac{1}{2}e^{-\tilde{\lambda}}\tilde{\zeta}_{a\pm}\tilde{\zeta}_{\mp}^{a}\right\}.
\end{equation}

and subject to first-class constraints
\begin{multline}
\\ \tilde{\Psi}_{Ba}(v^{a})=0,\\
 \mathcal{D}_{B}(\kappa)=0.\\
\end{multline}

As was initially stated, these constraints indicate that this theory possesses invariance under diffeomorphisms and Weyl transformations that preserve the preferred double null foliation. 
\section{Conclusions and Discussion}

In this essay, I have presented a gauge theory which possesses Diffeomorphism
and Weyl invariance that preserves the double null foliation and is
dual, and therefore equivalent to, general relativity adapted to that
foliation. As initially advertised, this theory propagates the same
degrees of freedom the general relativity does. In the body of the
essay, I have shown why this has to be the case by construction
of the symmetry trading algorithm which was employed in constructing
the dual theory. To be more precise, the dual theory is equivalent
to general relativity in the double null formalism where a preferred
parameterisation of the null hyper surfaces has been picked out. This
was to be expected as the symmetry trading algorithm required a gauge
fixing of at least one of the original theory's constraints.

There are some caveats to be made for this construction. In the construction
of the dual theory presented in the previous section, I haven't included
boundary terms in the original Lagrangian (\ref{eq:cdg}). In the
ADM formalism of Hamiltonian general relativity, such boundary terms
are known to give rise to boundary contributions to the Hamiltonian,
and are required to make the variational principle well defined. This
would be their purpose in the above framework as well. However these
terms wouldn't change the equations of motion and so formally, if
we repeat the above construction in the space of solutions to the
field equations where we have been careful with the boundary terms,
we can at most find some $\delta$ exact modification of the symplectic
potential but no obstruction to the construction itself should arise.

Another issue of concern is the formation of caustics and crossings
between different null hyper-surfaces. Caustics can be thought of
as points where neighbouring generators meet, but more precisely they
are points where the generators actually focus. These pathologies
are troublesome for they render the double null foliation unreliable.
In the body of the essay, this was implicitly taken into account
in the subsection of section 1 pertaining to the solution to the characteristic
initial value problem where the claim was that the solution to the
field equations are determined locally in the vicinity of two intersecting
null hyper surfaces, while no claims were made about global properties
of the solution. So this construction is safe so long as we are miserly
in treating only local regions of spacetime perhaps bounded by null
surfaces. Nevertheless, we can adopt a parameterisation which allows
us to explicitly truncate the null surfaces before caustics form.
The parameter which is sensitive towards the formation of caustic
is the so called `area parameter' $v$ defined as $v^{2}=\frac{\rho}{\rho_{0}},$
where $\rho_{0}$ is the area density defined on some initial space
like surface $\mathcal{S}_{o}$. This was employed in (\cite{miguel})
to construct the symplectic form for general relativity in terms of
free null initial data. On either null hyper surface $\Sigma_{0}$
and $\Sigma_{1}$ are considered demarcated by the two space like
surfaces where the area parameter vanishes $\mathcal{S}_{L}$, $\mathcal{S}_{R}$.
This is how caustics and crossing can be avoided. %

There is of course the question of the utility of what has been constructed
in this essay. In principle, the dual theory can be considered attractive
from the perspective of quantisation, if one were to find a set of
observables consistent with the gauge invariances of the theory. Then
the symplectic form can be used to deduce what the Peirels brackets
between these observables is which might be amenable to quantisation.
At this stage, this task seems very daunting and technically very
difficult considering how little is known about observables in the
double null framework. Another interesting application could be to
study issues relating to graviton scattering at null infinity in asymptotically
flat spaces. After all, past and future null infinity are null surfaces
which intersect at a Christodoulou-Kleinerman space, so this seems
to suggest that the problem might be amenable to analysis under the
lens of the theory dealt with in the body of this article. The most
direct application this theory can find in a physical problem is its
potential applicability towards the proof of the quantum Bousso bound
(see (\cite{BOUS}) for details) in the case of a conformally coupled
scalar field. This is because light-sheets are very easily accommodated
into this framework as null surfaces where the space like cross sections
have nowhere increasing expansion and the Linking Theory construction
and the Linking Theory is the same as the null phase space description
of the theory of a conformally coupled scalar field. What I mean by
potential applicability in the previous sentence is that the Linking
Theory provides the tools necessary to study this situation. This
shall be an area of future investigation.
\section{Acknowledgements}
I would like to sincerely thank my supervisor, Lee Smolin for his guidance and encouragement. I would like to thank Henrique Gomes and Flavio Mercati for carefully reading the draft of this essay and offering critical comments and suggestions. I am also grateful to Sean Gryb, Gabriel Herczeg, Tim Koslowski, Michael Reisenberger, Richard Epp, Yasha Neiman, Barak Shoshany, Illan Halpern and all my classmates in the PSI class of 2015 for stimulating discussions. This work was supported by the Perimeter Scholars International program. 
\section{Glossary of Symbols}

\begin{tabular}{|c|c|}
\hline 
Symbol & Definition\tabularnewline
\hline 
\hline 
$\mathcal{B},\mathcal{M}$ & Time-like and space-like three surfaces\tabularnewline
\hline 
$\Sigma_{o},\Sigma_{1}$ & Null three surfaces (in-coming and out-going)\tabularnewline
\hline 
$\mathcal{S}$ & Space-like two surface\tabularnewline
\hline 
$T_{p}\mathcal{S}^{\perp}$ & Space of normal vectors to $\mathcal{S}$ at a point $p$\tabularnewline
\hline 
$m_{\mu},s_{\mu},l_{\mu},k_{\mu}$ & Time-like, space-like and null normals to $\mathcal{S}$\tabularnewline
\hline 
$g_{\mu\nu}$ & Space-time metric\tabularnewline
\hline 
$\gamma_{ab},\Pi_{A}^{ab}$ & Intrinsic metric on the space-like two surfaces, and conjugate momentum\tabularnewline
\hline 
$u^{A}$ & Pair of parameters on the null surfaces\tabularnewline
\hline 
$\lambda$ & Scalar measuring the deviation of $n_{o},n_{1}$ from being parallel\tabularnewline
\hline 
$n_{\mu A}$ & Normal vector to a leaf of the double null foliation\tabularnewline
\hline 
$\varsigma^{AB}$ & Metric to raise and lower null direction indices\tabularnewline
\hline 
$\nu_{A}$ & In-affinity parameter\tabularnewline
\hline 
$\vartheta_{(b)}^{(a)},\mathcal{R}_{(b)}^{(a)}$ & Spin connection and curvature two-form on $\mathcal{S}_{\tau}$\tabularnewline
\hline 
$\partial_{A}$ & $\equiv\perp\mathcal{L}_{n_{A}^{a}}$\tabularnewline
\hline 
$K_{Aab}$ & Extrinsic curvature tensor of $\mathcal{S}_{\tau}$\tabularnewline
\hline 
$\sigma_{Aab},\theta_{A}$ & Shear tensor and expansion scalar of the null congruence\tabularnewline
\hline 
$\omega_{a},\omega_{\pm a},\zeta_{\pm a}$ & Normalized twist vector and linearly related quantities\tabularnewline
\hline 
$\varphi,\Pi_{\varphi A}$ & Scalar field and its momentum conjugate\tabularnewline
\hline 
$\mathcal{P}$ & Phase Space\tabularnewline
\hline 
$\Theta,J$ & (Pre-)Symplectic potential and current \tabularnewline
\hline 
$\Omega$ & (Pre-)Symplectic form\tabularnewline
\hline 
$X_{(\cdot)}$ & Hamiltonian vector field \tabularnewline
\hline 
$\{\cdot,\cdot\},\{\cdot,\cdot\}^{*}$ & Poisson brackets and Dirac brackets\tabularnewline
\hline 
$M_{IJ}$ & Second class constraint matrix\tabularnewline
\hline 
$\Psi_{Bo},\Psi_{Ba}$ & Constraints generating normal deformations and tangential diffeomorphisms
to $\mathcal{S_{\tau}}$\tabularnewline
\hline 
$\mathcal{D}_{A}(\kappa)$ & Constraint generating Weyl transformations restricted to $\mathcal{S}_{\tau}$
in dual theory\tabularnewline
\hline 
$\gamma_{o ab}$ & Fixed unimodular reference metric. \tabularnewline
\hline
$h_{ab}$ & Conformal two metric.\tabularnewline 
\hline
$\rho$& Densitized conformal factor\tabularnewline
\hline
$\varrho$& Exponential of $\rho$\tabularnewline
\hline
$\eta$&  Square of $\varrho$\tabularnewline
\hline
\end{tabular}

\end{document}